# Design Elements that Promote the Use of Fake Website-Detection Tools

**Fatemeh "Mariam" Zahedi**
University of Wisconsin-Milwaukee
zahedi@uwm.edu

**Ahmed Abbasi**
University of Virginia
abbasi@comm.virginia.edu

**Yan Chen**
University of Wisconsin-Milwaukee
yanchen@uwm.edu

**ABSTRACT**

Fake websites have emerged as a major source of online fraud, accounting for billions of dollars of loss by Internet users. We explore the process by which salient design elements could increase the use of protective tools, thus reducing the success rate of fake websites. Using the protection motivation theory, we conceptualize a model to investigate how salient design elements of detection tools could influence users' perceptions of the tools, efficacy in dealing with threats, and use of such tools. The research method was a controlled lab experiment with a novel and extensive experimental design and protocol. We found that trust in the detector is the pivotal coping mechanism in dealing with security threats and is a major conduit for transforming salient design elements into increased use. We also found that design elements have profound and unexpected impacts on self-efficacy. The significant theoretical and empirical implications of findings are discussed.

**INTRODUCTION**

Fake websites generate billions of dollars in fraudulent revenue by exploiting human vulnerabilities in online settings and are estimated to comprise nearly 20% of the Web (Abbasi et al. 2010; Gyongyi and Garcia-Molina 2005). Fake websites are a type of semantic attack in which attackers use meaningful content to semantically exploit weaknesses in human nature. Semantic attacks focus on "targeting the people" instead of exploiting hardware and software vulnerabilities as done by many other Internet security attacks, such as viruses, denial of service, and malware (Schneier 2000). Users play security-critical roles in that security failures could result not only from attacks but also from unpredictable human behaviors (Cranor 2008). The use of detection tools is the most important objective in the design of such tools since a tool that is turned off has no value in protecting users. To this end, we use a user-centered approach in which the relations between the critical design elements and actual use are investigated.

There has been little research on how salient design elements of protective IT artifacts could influence users' security perceptions and actual use. Previous studies (e.g. Wu et al. 2006) have been exploratory, and provided few theoretical insights regarding individuals' reactions. This paper is among the first to investigate the critical design elements of detection tools that could influence people's use of such tools by asking the specific research questions. What are the salient design elements of fake website detection tools that most impact users' security perceptions and promote use? What is the process by which the design elements alter users' behaviors?

In formulating the conceptual model to address the research questions, we draw on the protection motivation theory (PMT) (Rogers 1983). A novel experimental design, with extensive stimuli development using carefully identified spoofed and concocted websites, guided the data collection for this study. Our research uncovers the process by which the salient design elements of fake website detection tools enhance users' coping mechanism and increase the actual use of such tools.

This paper makes important and novel theoretical and empirical contributions. Our work shows how detection tools' design elements must be enhanced and marketed to promote their use. We found trust in the detector is the pivotal coping mechanism in dealing with security threats posed by fake websites and is a major conduit for transforming users' perceptions of design elements into their increased use of such tools. Elements that boost users' trust are critical factors in the design of detection tools. We also found that design elements have profound and unexpected impacts on users' self-efficacy, which indicates the potential presence of an ego-enhancing role for protective IT artifacts. Accuracy emerged as the most important design element that impacted trust in the tool, and its actual use. Our work shows how this primary feature of the tool operates on users' psychological mechanisms to change their perceptions and behaviors.

**THEORETICAL BACKGROUND AND MODEL DEVELOPMENT**

To identify the salient design elements for fake website detection tools and understand user reactions to such tools, we need to look into users' cognitive process of detecting deceptions when such tools are in the place.

According to the fraud deception theory (FDT), a deception involves two parties with conflicting interests, a deceiver and a target. The deceiver uses deceptive tactics to manipulate and misrepresent cues of a situation which depart from the truth, to induce a misjudgment by the target. The target, therefore will behave in accordance with the deceiver's manipulations and misrepresentations (Johnson et al. 2001). To successfully detect deceptions, individuals need to detect the inconsistencies between the cues manipulated and the truth. The FDT suggests a fraud detection method based on the competence model of successful detections that can assist an individual's cognitive process of fraud detection. Effective fake detection tools need to facilitate individual's cognitive process in arousing suspicion(s) about abnormalities, generating and evaluating hypotheses on deceptions, and reaching a conclusion on the legitimacy of a site (Johnson et al. 2001). Two broad categories of design element can facilitate such cognitive process, performance elements and user interface elements of detection tools. Performance elements play a critical role in activating users' fraud detection cognitive process before they fall into deceiver's manipulations; user interface elements communicate the findings of the tool and help users detect manipulated and misrepresented cues and then heed the tool's warnings. This paper is part of a larger federally-funded research project that investigates both performance design and user-interface design of fake website-detection tools. In this paper, we report on the performance-related design elements of such tools. (The research on the user-interface elements of detection tools is currently underway.) Based on the literature and the FDT theory, we have identified four categories of salient design elements, detector' accuracy and run time, cost of detector error, type of threat (spoofed and concocted), and domain.

**Self-Protective Behavior Research**

There are theories for explaining how individuals can be motivated to protect themselves from harm. Protection motivation theory (PMT) (Rogers 1983) is the most well-known in security research and has demonstrated significant explanatory power in predicting security behaviors (e.g. Anderson and Agarwal 2010; Chen and Zahedi 2009; Johnson and Warkentin 2010). PMT posits that humans' protective behaviors involve two cognitive processes—threat appraisal and coping appraisal. The principal variables in the threat appraisal process are the perceived susceptibility to the threat (a perception about the extent of vulnerability to the threat) and the perceived severity of the threat (a perception about the magnitude of possible harm of the threat if no countermeasures are taken). The primary constructs in the coping appraisal process are response efficacy (a belief in the effectiveness of the countermeasure), self-efficacy (a belief in one's own ability in taking the countermeasure), and perceived cost of adoption (a perception about the amount of effort, time, and money needed for adopting the countermeasure) (Rogers 1983). PMT proposes that when individuals appraise the threat by assessing the susceptibility to and severity of the threat, and are confident in their coping ability (in terms of response efficacy, self-efficacy, and coping cost), they tend to take protective action against the threat.

**Model and Hypotheses Development**

In order to investigate the impact of the salient design elements, we propose the detection tool impact (DTI) model, as shown in Figure 1.

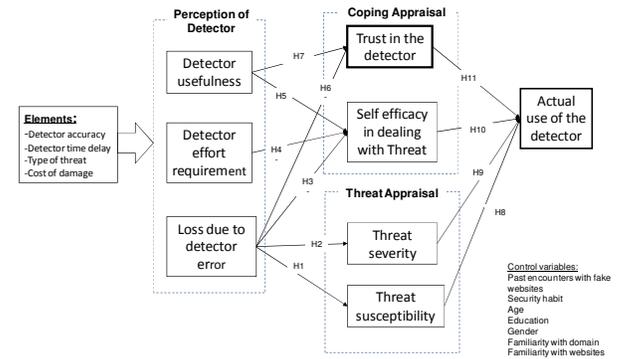

**Figure 1. Detector Tool Impact (DTI) Model**

This is a user-centric assessment of how the detection tool design could change the usage of the system. In this model, the manipulations of design elements form users' perceptions of the detector, which in turn impact users' cognitive processes—threat appraisal and coping appraisal and then actual use of the tool. The arguments for model conceptualization were not included here due to the space limitation. The model hypotheses are reported below.

H1. Users' perceived loss due to detector error is positively associated with their perceived threat susceptibility.
H2. Users' perceived loss due to detector error is positively associated with their perceived threat severity.
H3. Perceived loss due to detector error is negatively associated with users' self-efficacy in dealing with fake website threats.
H4. Detector effort requirement is negatively associated with users' self-efficacy in dealing with fake website threats.
H5. Detector usefulness is positively associated with users' self-efficacy in dealing with fake website threats.
H6. Perceived loss due to detector error will have a negative effect on users' trust in the detector.
H7. Detector usefulness is positively associated with users' trust in the detector.
H8. Users' perceived threat susceptibility is positively associated with the extent of their detector use.
H9. Users' perceived threat severity is positively associated with the extent of their detector use.
H10. Users' perceived self-efficacy is positively associated with the extent of their detector use.
H11. Users' trust in the detector is positively associated with the extent of their detector use.

## RESEARCH METHODOLOGY

The research methodology was controlled lab experiment. We chose this methodology in order to examine the actual behaviors of individuals in using the detector. The experimental design was a 2 x 2 x 2 x 2=16 full-factorial design. The detection tool (DT) stimuli varied in design elements: accuracy, run time, type of threat, and loss due to detection error. Each design element had two levels: the DT accuracy was high vs. low (90% vs. 60%), the detection run time was fast vs. slow (1 vs. 4 seconds), the type of threat was either spoofed or concocted, the loss due to the detection error was high vs. low ($10 vs. $1). The values of all manipulations were determined by the suggestions from literature (e.g. Abbasi et al. 2010).

The experimental protocol was to mimic real conditions of use. We chose the sensitive context of online pharmacy. An inventory of 15 spoofed, 15 concocted, and 15 legitimate online pharmacies was identified. A detection tool simulated the performance of fake website-detection based on one of the 16 possible designs. The designs were randomly assigned to the participants. The participants were randomly assigned 5 legitimate and 5 fake (either spoofed or concocted) websites. The experimental task was to buy an over-the-counter drug with a value of about $30 (Rogaine, a hair regrowth product, for grandpa). This product was chosen because it is relevant and familiar. For each website, the participants had to decide if they would visit the website, and once on the website, the participants had to decide if they would explore the website to find the product and, once found, if they would buy the product. Incentives and/or disincentives are also important factors to motivate human beings to take the proper security actions when they are in the security loop (Cranor 2008). The participants were therefore awarded based on their performance. The final performance score for each participant was computed based on all their decisions regarding 10 assigned websites. Payments were made in cash (gift card) or course credits based on each participant's preference. The participants took an online survey before and after the experiment.

All the scales for the study were also adopted from literature and modified for the current study. All items were converted to have semantic differential scales to ensure content validity and reduction of the threat of common method variance (Podsakoff et al. 2003). The instrument and corresponding references were not included due to space limitations.

The experiment protocol and the instrument were pretested and pilot-tested. 437 participants from students in undergraduate and graduate programs in a large Midwestern university participated in the experiment. Average age of participants was 22.4 years, with 64% male and 36% female.

## ANALYSIS AND RESULTS

In the post-experiment survey, participants were asked to assess the detection time and accuracy and the cost of making one wrong decision, which we manipulated during the experiment, based on what they had experienced in the experiment. The results of the ANOVA tests supported the success of manipulation.

We carried out exploratory factor analyses (EFA) to check the convergent and discriminant validity of the constructs. All measurement items emerged and correctly loaded on the corresponding constructs. All item loadings are greater than 0.70, and no cross loadings are greater than 0.40. By comparing the square root of the AVE for each construct with its correlations with all other constructs, we checked for further evidence of the discriminant validity of the constructs. The square root of AVE for each construct was greater than the correlation values with other constructs. All together, the convergent and discriminant validity were supported.

### Measurement Model

The measurement model was estimated by using the mean-adjusted maximum likelihood (MLM) method in Mplus. MLM adjusts the estimation for the non-normality in data. As shown in Table 1, all fit indices were better than recommended thresholds, indicating a good fit for the measurement model.

| Fit Index | Measurement | DTI | Threshold |
|---|---|---|---|
| Normed $\chi 2$ | 1.31 | 1.61 | <3.0 |
| CFI | 0.992 | 0.973 | >0.90 |
| TLI | 0.990 | 0.970 | >0.90 |
| RMSEA | 0.027 | 0.037 | <0.06 |
| SRMR | 0.028 | 0.072 | <0.10 |

**Table 1. Measurement and Casual Model Fit Indexes**

### Model Estimation

We used the MLM method in Mplus to estimate the DTI model. As shown in Table 1, all the fit indices of the DTI model were better than the recommended thresholds, indicating good model fit and supporting for our theoretical model. Of the 11 hypotheses in the DTI model, 9 were statistically significant (Hypotheses H8 and H9 were not supported), confirming our theoretical conceptualization.

## DISCUSSION

The experiment design and execution proved to be successful in manipulating participants' perceptions about the detection tool, thus allowing the estimation of the conceptual model in order to examine how the design elements could change users' perceptions and behaviors regarding detecting fake websites. Our results indicated that users form their perception of the detector's usefulness based on its accuracy, which, in turn, plays a major role in forming users' trust in the detector. Run time and cost due to detector error have far smaller roles in users' perception of the detector's attributes.

Trust in the detector emerged as the single most pivotal factor in linking detection tools' design elements (H6 and H7) to use behaviors (H11). Positive perception about detectors (usefulness) arises from detectors' accuracy and has the single most profound influence on trust (path coefficient of 0.89, p<0.001), whereas loss due to detector error has a small negative influence (-0.07, p<0.01). This shows that forming and promoting positive perceptions about detectors are critical in developing users' trust, which, in turn, promotes use. Given the fact that people generally are not good at detecting deceptions (Biros et al. 2002), we believe users can substantially benefit from developing trust in detectors that they find useful. Our findings further support the importance of giving proper consideration to design elements (such as detection accuracy) that "calibrate" human trust in protective IT artifacts (Parasuraman and Miller 2004) to promote trust in and use of protective IT artifacts.

Loss due to detector error has a small negative influence on trust in the detector. Its most damaging impacts are on users' self-efficacy and threat appraisal, particularly on threat susceptibility. These findings indicate that the detector's negative attributes, while perceived as contributing to threat and reducing users' self-efficacy in taking security countermeasures to deal with the threat, may not play a significant role in user behavior so long as the tool is perceived as useful and, hence, trustworthy due to its positive design elements.

Our results highlighted and supported numerous previous findings that self-efficacy is an important salient construct. However, most IS studies treat self-efficacy as an exogenous variable and do not examine forces contributing to its change, either positively or negatively. Our work shows that detectors' positive attributes increase self-efficacy (H5), and detectors' negative attributes reduce self-efficacy (H3). This is an interesting finding since it indicates that users make a connection between the "ability" of an IT artifact and their own ability. When perceived loss due to detector error is high, users lower their perception of self-efficacy (H3); when perceived detector usefulness is high, they increase their perception of self-efficacy (H5). This is a novel finding since it shows that the detector has the potential to merge with users' ego and to become part of their self-perception—"I have more ability since I have a more powerful tool." This has the potential to blur the boundary of the self and the IT protective artifact.

This finding becomes even more interesting when we consider the unexpected result for effort requirement. The detector's effort requirement is another negative attribute. Its impact was not as hypothesized in H4. We had hypothesized a negative impact on self-efficacy, whereas its impact turned out to be significant but positive. The inconvenience of waiting for a few seconds and the cognitive effort of reading and deciding about the warning message were not an issue for users in our sample. Instead, the effort provided users with a sense of increased self-efficacy. It seems that seeing the detector in operation gave users a higher sense of control and power in dealing with the fake website threat. Again, the boundary of self and the detector becomes blurred in this interpretation since the effort is accepted as an ego-enhancing process in the fight against the fraud perpetrated by fake websites. This also explains why trust in the detector plays such an important role. Trust by definition indicates a close relationship in which the trustor is willing to become vulnerable to the trustee's actions and accept the trustee's actions without verification. It raises the question of whether protective IT artifacts are perceived as extensions of users' self and own ability. Does the artifact become the "Iron Man's armor" to give him super power?

Finally, we examined the DTI model with two types of fake website threat: spoofed and concocted. The findings of the estimated model remain the same in both types of threat, indicating the generalizeability of the DTI model.

**IMPLICATIONS, LIMITATIONS, AND FUTURE DIRECTIONS**

This paper makes a number of novel contributions to theory and practice. First, using the theoretical lens in designing IT protective artifacts opens a new avenue of research. Second, this study addressed the call for research in the relationship between trust and the IT artifact in general and the relationship between trust and the detection system assisting users to detect online deceptions in particular (Gefen et al. 2008). Our investigation uncovered the pivotal role of trust in the artifact as the conduit between design elements and their eventual impacts. When designing protective IT artifacts, designers should give proper consideration to design elements regarding the trust "calibration" process (Parasuraman and Miller 2004). Third, our findings uncovered the ways that users' perceptions of design elements and artifact attributes influence their self-efficacy by either enhancing or diminishing their sense of personal ability to cope with the threat of fake websites. Users may view protective IT artifacts as an extension of their self, thus reinforcing the need to combine the design science approach with behavioral theories to fit protective IT artifacts to individuals' psychology in order to promote users' trust and use. Fourth, this study contributes to design science research by proposing and empirically testing the DTI model by which design science scholars can test and evaluate various design elements of protective IT artifacts. Finally, our findings indicated that the main focus of research in designing protective IT artifacts should be on coping mechanisms and not on the appraisal of threat and fear. This increases the parsimony of model conceptualization for designing protective IT artifacts.

This study has limitations. We collected our data for one context—online pharmacies. Therefore, our results should be interpreted within this context. Our participants interacted with the detection tool stimuli that were not embedded in an Internet browser or running as a real-time system. This could be considered a limitation. However, our stimuli closely imitated main features of existing detection tools while eliminating specifics of a brand name. Thus, participants had a unified experiment platform and, consequently, variances due to participants' varying experiences and knowledge of specific tools were removed. Further, this study was conducted with undergraduate and graduate students. Although undergraduate and graduate students represent a large proportion of Internet users, care must be taken in generalizing our findings to other populations.

This paper is part of a larger, federally-funded project in this area that involves the investigation of user-interface elements, multiple contexts, and personalization through intelligent user interface. Additionally, this paper has proposed a number of avenues for building theories that combine the strength of design science and behavioral science in creating compelling personalized protective IT artifacts. Another direction of future research is the use of the DTI model in examining the design elements for other protective IT artifacts. Given the strong empirical support for the relationship between trust in the detector, it might be promising for future research to study how trust in protective IT artifacts is extended to the Internet. Finally, other Internet usage domains, particularly hedonic domains such as online games and social networking, could be explored in future research. It is possible that the impacts of the detector in hedonic domains will be different from those in utilitarian domains such as the online pharmacy employed in this study.

## ACKNOWLEDGMENTS

This material is based upon work supported by the National Science Foundation under Grant No. CNS-1049497.